\documentclass[showpacs,preprintnumbers,amsmath,amssymb]{revtex4}
\usepackage{graphicx}% Include figure files
\usepackage{dcolumn}% Align table columns on decimal point
\usepackage{bm}% bold math

\begin{document}

%\preprint{APS/123-QED}

\title{Spin-orbital coupling effect on Josephson current through a superconductor heterojunction}
\author{Z.H. Yang$^{1}$, Y.H. Yang$^{1}$, J. Wang$^{1,2}$, and K.S. Chan$^{2}$}
\address{$^{1}$Department of Physics, Southeast University, Nanjing 210096, China}
\address{$^{2}$Department of Physics and Materials Science, City University of Hong Kong, Tat Chee
Avenue, Kowloon, Hong Kong, China}

\date{\today}

\begin{abstract}
We study spin-orbital coupling effect on the Josephson current
through a superconductor (SC) heterojunction, consisting of two
s-wave superconductors and a two-dimensional electron gas (2DEG)
layer between them. The Rashba-type (RSOC) and/or Dresselhaus-type
(DSOC) of spin-orbital coupling are considered in the 2DEG region.
By using the lattice Bogoliubov-de Gennes equation and the Keldysh
formalism, we calculate the DC supercurrent flowing through the
junction and find that the critical current $I_c$ exhibits a damped
oscillation with both the strength of SOC and the layer length of
2DEG; especially, the strength ratio between RSOC and DSOC can also
induce switching between the $0$ state and the $\pi$ state of the
SC/2DEG/SC junction as well. This $0$-$\pi$ transition results from
the fact that SOC in a two-dimension system can lead to a
pseudo-magnetic effect on the flowing electrons like the effect of a
ferromagnet, since the time reversal symmetry of the system has
already been broken by two SC leads with different macroscopic
phases.
\end{abstract}

\maketitle

The issue of Josephson current sign reversal in a
superconductor/ferromaget/superconductor (SC/FM/SC) heterojunction
has recently drawn a great deal of attention\cite{1,2,3} due to its
experimental observation and potential applications to spintronics
and quantum computing.\cite{4,5,6,7,8} The physical origin of the
Josephson effect is the breakdown of time reversal symmetry (TRS) in
SC/SC junctions due to the SC macroscopic phase difference. Between
the two SC leads in the junction is either an insulator or a "weak
link" such as normal metal (NM), semiconductor, and so on. The DC
Josephson effect can be understood by the Andreev reflection
processes of quasiparticles\cite{9} with energy smaller than the
superconducting energy gap, an electron impinging on one of the
interfaces is Andreev reflected and converted into a hole moving in
the opposite direction, thus generating a Cooper pair in an SC; this
hole is consequently Andreev reflected at the second interface and
is converted back to an electron, leading to the destruction of a
Cooper pair in the other SC. As a result of this cycle, a pair of
correlated electrons are transferred from one SC to another,
creating a supercurrent flow across the junction.\cite{1}

When an FM is inserted between the two SC leads, the
current-carrying Andreev bound states are spin-splitted and shifted
in an oscillatory way. FM favors to align the electron spin due to
the FM exchange interaction, whereas the ordinary spin singlet
Cooper pair consists of two spin-antiparallel electrons; thus FM
tends to destroy the superconductivity and the Josephson current in
SC/FM/SC junctions would be depressed compared with SC/NM/SC
junctions. Furthermore, the Cooper pair is associated with a nonzero
momentum due to spin splitting from the FM exchange interaction and
the Josephson current exhibits a damped oscillating behavior with
the FM layer length or the FM exchange strength, i.e., the $0$-$\pi$
transition.\cite{10,11,12,13,14,15,16,17,18,19,20,21,22,23} The
$\pi$-state of the SC/FM/SC junction is the Josephson current
flowing in the direction opposite to the phase difference between
the two SCs. To realize systematically the $0$-$\pi$ state switching
in a single SC/FM/SC junction, it needs to change either the length
of the FM layer or the FM exchange strength; however, as a matter of
fact, it is very difficult to do this in an experiment. Since the
tunability of a system is of importance for experimental
observation, some alternatives have been proposed; for example, the
SC/FM/FM/SC junction with noncollinear magnetizations has been
investigated by Pajovi\'{c} \emph{et al.}\cite{24} and the $0$-$\pi$
transition can be found by modulating the relative direction of the
FM moments. Dolcini and Giazotto\cite{25} proposed to use
Aharonov-Bohm interferometry to realize a fully controllable
Josephson junction.

The Josephson current flowing through a two-dimensional electron gas
(2DEG) region with Rashba (RSOC) and/or Dresselhaus (DSOC) spin
orbital coupling has also been studied, because the pseudomagnetic
field from SOC may lead to the same effect on the Josephson current
as FM.\cite{26,27,28,29} Several studies\cite{26,27} have shown that
in the one-dimension case the SOC can hardly exert any effect unless
the Zeeman splitting from an external magnetic field is included. It
was generally argued that SOC does not break the TRS of the system
and cannot function like an FM. However, Dell'Anna \emph{et
al.}\cite{29} argued that TRS has already been broken by the
supercurrent and found that SOC can have a huge effect on the
Josephson current in a quantum dot system, although they did not
find the $0$-$\pi$ transition behavior since only two electron
levels are considered in the quantum dot.

In this work, we restudy the SOC effect on the DC Josephson current
through a SC/2DEG/SC junction in the clean limit and with the
multilevels in the 2DEG being taken into account. When TRS of the
system is broken by the two SCs with different macroscopic phases,
the pseudomagnetic field from SOC should cause an effect (hereafter
referred to as pseudo-magnetic effect) on the electron transport
property, resembling the exchange field effect in a FM; however, for
1D transport, the pseudo-magnetic effect from SOC should
disappear.\cite{30} Based on this analysis, we calculate the
Josephson current flowing in a SC/2DEG/SC junction by using the
discrete BdG equation and Keldysh Green's functions. It is shown
that the Josephson current exhibits a damped oscillation as a
function of the strength of SOC, the length of the 2DEG layer, and
the strength ratio between RSOC/DSOC. This $0$-$\pi$ transition is
the same as that found in SC/FM/SC junctions. However, the $0$-$\pi$
transition induced by SOC is much easier to be realized in
experiment since the strength of SOC can be modulated to a large
extent by an external electric field perpendicular to the plane of
2DEG.\cite{31}

We consider a clean SC/2DEG/SC heterojunction, consisting of two
semi-infinite SC leads and a 2DEG layer with a length $d$ between
them, as shown in Fig.~1. The interfaces between 2DEG and SCs are
set at $x=0$ and $x=d$ and the current flows along the
$x$-direction. In the 2DEG layer both RSOC and DSOC are considered,
which come from the structure asymmetry of the material\cite{32,33}
and are expressed as
\begin{equation}
{\cal{H}}_{SOC}=\frac{\alpha}{\hbar}(\sigma_y p_x-\sigma_x p_y)+
\frac{\beta}{\hbar}(\sigma_x p_x-\sigma_y p_y),
\end{equation}
where $\alpha$ and $\beta$ are the coupling constants of RSOC and
DSOC, repectively, and RSOC can be directly modulated by an external
electric field perpendicular to the 2DEG plane,\cite{31} $p_x$ and
$p_y$ are the two components of the momentum operator $\textbf{p}$,
$\sigma_x$ and $\sigma_y$ are the Pauli matrices. The difference
between these two SOCs is the different directions of the
pseudomagnetic fields from SOC for the same electron momentum. By
introducing a ratio angle $\theta$, the equation above can be
rewritten as
\begin{equation}
{\cal{H}}_{SOC}=\frac{\gamma}{\hbar}(\tilde\sigma_y
p_x-\tilde\sigma_x p_y).
\end{equation}
Here $\gamma=\sqrt{\alpha^2+\beta^2}$, $\alpha=\gamma\cos\theta$,
and $\beta=\gamma\sin\theta$, $\tilde\sigma_x=
\left(%
\begin{array}{cc}
  0 & e^{-i\theta} \\
  e^{i\theta} & 0 \\
\end{array}%
\right)$, and  $\tilde\sigma_y=\left(%
\begin{array}{cc}
  0             & -ie^{i\theta} \\
  ie^{-i\theta} & 0              \\
\end{array}%
\right)$. The angle $\theta$ denotes not only the strength ratio
between RSOC and DSOC, but also the pseudomagnetic field direction
for the mixture of RSOC and DSOC, which is similar to the rotation
angle of the spin axis,\cite{30} as can be seen in the renormalized
$\tilde\sigma_x$ and $\tilde\sigma_y$.

The two SC leads are described in the framework of standard BCS
formalism with a macroscopic phase difference $\phi$. In the absence
of external applied field, the mean-field Hamiltonian of the
SC/2DEG/SC junction reads\cite{34}
\begin{mathletters}
\begin{eqnarray}
{\cal{H}}={\cal{H}}_L+{\cal{H}}_R+{\cal{H}}_{2DEG}+{\cal{H}}_T,
\end{eqnarray}
\begin{eqnarray}
{\cal{H}}_{L(R)}=\sum_{ij\sigma}(\varepsilon
_{ij\sigma}-\mu)C_{ij\sigma}^{\dagger}C_{ij\sigma}-t\sum_{ij\sigma}
(C_{i+1,j\sigma}^{\dagger}C_{ij\sigma}+C_{i,j+1,\sigma}^{\dagger}C_{ij\sigma}+c.c.)
\nonumber \\
+\sum_{ij}(\Delta_{L(R)}C_{ij\uparrow}^{\dagger}C_{ij\downarrow}^{\dagger}+
\Delta^{*}_{L(R)}C_{ij\downarrow}C_{ij\uparrow}),
\end{eqnarray}
\begin{eqnarray}
{\cal H}_{2DEG}=\sum\limits_{ij\sigma}(\varepsilon
_{ij\sigma}-\mu)C_{ij\sigma}^{\dagger}
C_{ij\sigma}-t\sum\limits_{ij\sigma}\{C_{i+1,j\sigma}^{\dagger}C_{ij\sigma}
+C_{i,j+1,\sigma}^{\dagger}C_{ij\sigma}+c.c\}  \nonumber \\
-t_{so}\sum\limits_{ij\sigma\sigma^{\prime}}\{C_{i+1,j\sigma}^{\dagger}
({\it{i}}\tilde\sigma_{y})_{\sigma\sigma^{\prime}}C_{ij\sigma^{\prime}}-
C_{i,j+1,\sigma}^{\dagger}({\it{i}}\tilde\sigma_{x})_{\sigma\sigma^{\prime}}
C_{ij\sigma^{\prime}}+c.c.\},
\end{eqnarray}
\begin{eqnarray}
{\cal
H}_{T}=\sum_{j\sigma}(t_{L}'C_{j\sigma,L}^{\dagger}C_{j\sigma,2DEG}+
t_{R}'C_{j\sigma,R}^{\dagger}C_{j\sigma,2DEG}+c.c.).
\end{eqnarray}
\end{mathletters}
Here ${\cal{H}}_L$ and ${\cal{H}}_R$ are respectively the
discretized BdG equation of the left and right SC leads;
${\cal{H}}_{2DEG}$ is the Hamiltonian of the 2DEG region, i.e., a
free electron model with SOC included; ${\cal{H}}_T$ is the
tunneling Hamiltonian between the 2DEG and the left and right SC
leads. $C_{ij\sigma}^{\dagger}(C_{ij\sigma})$ is the creation
(annihilation) operator of an electron at site ($ij$) with spin
$\sigma=\uparrow\downarrow$, $\varepsilon_{ij\sigma}=4t$ is the
site-energy, $t={\hbar}^{2}/2ma^{2}$ is the hopping energy with the
lattice constant $a$ , and $t_{so}=\gamma/2a$ denotes the SOC
strength in the lattice representation.
$\Delta_{L(R)}=|\Delta_{L(R)}|e^{\pm i\phi/2}$ is the pair potential
in the left (right) SC lead with a macroscopic phase $\pm\phi/2$,
$\mu$ is the chemical potential which is constant everywhere in our
model due to zero applied bias in the junction. $t'_{L(R)}$ is the
hopping strength between the left (right) lead and 2DEG, which is
independent of spin so that no spin flip effect occurs when
quasiparticles tunnel through the interfaces; it can also represent
the strength of the interface barrier between the SC lead and the
2DEG region.

Although we consider a clean system without impurity, the
Hamiltonian above is not limited to this system and can describe
systems with other band structures as well as different pair
potential symmetry. Along the $y$ direction of the junction in
Fig.~1, the translational symmetry is preserved so that the electron
momentum $k_y$ is a good quantum number and summation in Eq.~3 over
$j$ sites can be transformed to the $k$-space, e.g.,
\begin{eqnarray}
{\cal
H}_{2DEG}=\sum_{i\sigma}(4t-\mu-2t\cos{k_ya})C^{\dagger}_{i\sigma}C_{i\sigma}
-\sum_{i\sigma}(tC^{\dagger}_{i+1\sigma}C_{i\sigma}+c.c)-\nonumber
\\
\sum_{i\sigma\sigma'}(t_{so}C^{\dagger}_{i+1\sigma}({\it{i}}\tilde\sigma_{y})C_{i\sigma'}+c.c.+
2t_{so}\sin{k_ya}C^{\dagger}_{i\sigma}(\tilde\sigma_{x})C_{i\sigma'}).
\end{eqnarray}
Other components of the Hamiltonian (Eq.~3) can be treated in the
same way, thus the numerical calculation will be reduced greatly
compared with a finite size (along the $y$-direction) system. We
focus on the supercurrent through the interface between the SC and
2DEG, thus, the current density operator is given by
$\hat{J}_L=e\frac{d\hat{N}_L}{dt}=\frac{1}{i\hbar}[\hat{N}_{L},H_{T}]$
where $\hat{N}_{L}$ is the electron operator in the left SC, and
after commutation the steady current density reads\cite{35}
\begin{equation}
J_{L}=\frac{e}{\hbar W}\int
\frac{dE}{2\pi}\sum_{k_y}\{t'_{L}G_{L,2DEG}^{<}(E,k_y)-
(t'_{L})^*G_{2DEG,L}^{<}(E,k_y)\}_{11+33}.
\end{equation}
$G^{<}(E,k_y)$ is the lesser Green's function defined in $4\times 4$
Nambu$\bigotimes$spin space,
$G^{<}_{i\sigma,j\sigma'}(t,t')=i\langle
C^{\dagger}_{j\sigma'}(t')C_{i\sigma}(t)\rangle$ with
$\langle\ldots\rangle$ denoting the quantum statistical average. $W$
is the transverse width of the junction, the subscripts $11+33$
stands for the summation over the spin-up and spin-down components
of the electron current, i.e., the $G^<$ (1,1) and (3,3) matrix
elements. Since we consider the DC Josephson effect and no bias is
applied on the system, the lesser Green's function could be obtained
by the equilibrium equation $G^<(E)=(G^a(E)-G^r(E))f(E)$ with $f(E)$
being the Fermi-Dirac distribution function, $G^{r(a)}$ is the
retarded (advanced) Green's function, which fulfills the Dyson
equation, e.g.,
\begin{equation}
G^{r(a)}_{L,2DEG}=g^{r(a)}_L\tilde{H}_{TL}G^{r(a)}_{2DEG},
\end{equation}
where $g^{r(a)}_L$ is the surface Green's function of the left SC
lead and $G^{r(a)}_{2DEG}$ is the coupled Green's function of the
2DEG, $\tilde{H}_{TL}$ is the left tunneling matrix of Eq.~(3d) in
the Nambu$\bigotimes$spin space. The retarded Green's function can
be calculated by direct matrix inversion
$G^r_{2DEG}=(E-\tilde{H}_{2DEG}-\Sigma_L^r-\Sigma_R^r)^{-1}$  or the
recursive method,\cite{36} where $\tilde{H}_{2DEG}$ is the 2DEG
Hamiltonian Eq.~(3b) in the Nambu$\bigotimes$spin space and
$\Sigma_{L(R)}^{r}$ is the self-energy from the left (right) SC
lead.

In the numerical calculation, the hopping energy $t$ is set as the
energy unit $t=1$, the chemical potential takes $\mu=2.0t$ and the
pair potentials in the two SC leads are assumed to be equal,
$|\Delta_{L(R)}|=0.01\mu$. The Josephson current is calculated at
zero temperature based on Eq.~5 and the current phase relations are
shown for different strength ratio angles $\theta$ in Fig.~2. The
hopping term $t_{L(R)}'$ stands for the interface transparency
between the SC leads and 2DEG; $t_{L}'=t_{R}'=t$ is set in the
calculation to denote full transparency of the interfaces so that
$I(\phi)$ deviates from the sinusoidal relation as shown in
Fig.~(2a). The $\theta=\pi/4$ case (dashed-line) exhibits the
coexistence of the $0$ and $\pi$ states, which suggests a $0$-$\pi$
transition of the junction  should  occur upon changing the angle
$\theta$ and other parameters. In Fig.~(2b), it is expected that
$I(\phi)$ can approximate a harmonic function for low interface
transparency $t_{L}'=t_{R}'=0.5t$.

In Fig~3, the critical Josephson current $I_c$ is presented as a
function of the angle $\theta$ for different SOC strength $t_{so}$
and 2DEG layer length $d$. The dips in these curves correspond to
the $0$-$\pi$ transitions and the oscillation period decreases with
the SOI strength. Changing $\theta$ is equivalent to modulating the
relative pseudomagnetic field direction, which is perpendicular to
the electron-momentum direction for RSOC while for DSOC, the
pseudomagnetization direction is complementary to RSOC by exchanging
the spin axis.\cite{30} Here the mixture of RSOC and DSOC is similar
to the SC/FM/FM/SC junction with noncollinear FM moments studied by
Pajovi\'{c} \emph{et al.},\cite{24} who found $I_c$ can oscillate
with the angle between the two magnetizations. In Fig.~4 and Fig.~5,
the critical current $I_c$ is also shown to exhibit the $0$-$\pi$
transition by varying either the SOC strength $t_{so}$ (Fig.~4) or
the 2DEG layer length (Fig.~5). The damped oscillation of $I_c$
suggests that the $\pi$-state of the junction stems from the
psuedo-magnetic effect of SOC, which is similar to an SC/FM/SC
junction in terms of physical origin.

The $\pi$ state of the SC/FM/SC junction results from the Cooper
pair (coherent electron and hole in FM) being associated with a
nonzero momentum due to the FM exchange interaction, which can lead
to an $I_c$ oscillating with the FM exchange strength as well as the
FM layer length. For the studied SC/2DEG/SC junction, SOC maintains
the TRS of the system itself on the whole because each electron sees
a pseudomagnetic field $B=\frac{\gamma}{\hbar}(\hat{z}\times
\textbf{p})$ from Eq.~2; the system does not have any
pseudo-magnetic effect, which resembles breaking the TRS, when a
summation is taken over all electrons. However, in the present case
the TRS of system has already been broken by the current, and the
SOC should have effects on the transport properties. By summing over
those electrons (momenta) contributing to the current, the
pseudo-magnetic effect of SOC appears; for example, the anomalous
tunneling magnetoresistance (ATMR) observed in experiments\cite{37}
is generally attributed to the SOC effect which is a pseudo-magnetic
effect on the TMR. When a Josephson current is flowing through the
SC/2DEG/SC junction,  SOC can affect it definitely and $I_c$ can
exhibit a damped oscillation like in an SC/FM/SC junction.

For one-dimensional SC/2DEG/SC junctions with SOC or only
perpendicular injection considered ($k_y=0$), one cannot obtain
$0$-$\pi$ transition as those shown in Fig.~(3-5). The SOC cannot
lead to a pseudo-magnetic effect in the 1D case, because in this
case the density of states is still spin-degenerate, unlike in the
2D case.\cite{30} Therefore, $k_{y}\neq 0$ is a prerequisite for SOC
exhibiting magnetic effect on electron transport and the summation
over all transverse modes $k_y$ does not mean the effect of SOC
would disappear, for the current is not simply an odd function of
$k_y$, at least in the clean limit. This is different from the
semiclassical average over $k_y$ in Ref.~28. Compared with SC/FM/SC
jucntions, the SOC strength as well the ratio angle are relatively
easy to modulate by an perpendicular electric field, this feature
may be of importance for experimental applications. For instance,
Nitta \emph{et al.} found the SOC constant can be modulated up to
50\% in InAs material.\cite{31}

In summary, we have studied SOC effect on the Josephson current
through a SC/2DEG/SC junction based on the lattice BdG equation and
Keldysh formalism. It was found that the critical Josephson current
$I_c$ can exhibit a damped oscillation with both the strength of
SOC, the layer length of the 2DEG, and the strength ratio angle
$\theta$. The $0$-$\pi$ transition, same as that found in SC/FM/SC
junctions, results from SOC induced pseudo-magnetic effect in a 2D
system when TRS of the system is broken by the supercurrent flow;
but for 1D junction the $\pi$ state is absent. The findings here can
be observed experimentally by modulating the strength of SOC with
external electric field.

ACKNOWLEDGEMENT  This work is supported by City University of Hong
Kong Strategic Research Grant (Project No. 7002029)and NSFC.
10704016.

\onecolumngrid

\newpage

\begin{figure}[htbp]
    \includegraphics[width=4in]{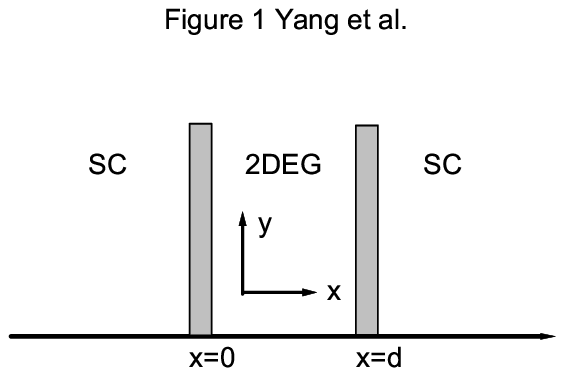}
    \caption{ The schematic of a two-terminal device in which two SC
leads connect with the 2DEG, the shadow parts are the two interface
barriers. Josephson current resulted from the macroscopic phase
difference of the two SCs flows through the 2DEG along the
$x$-direction.}
\end{figure}

\newpage

\begin{figure}[htbp]
    \includegraphics[width=4in]{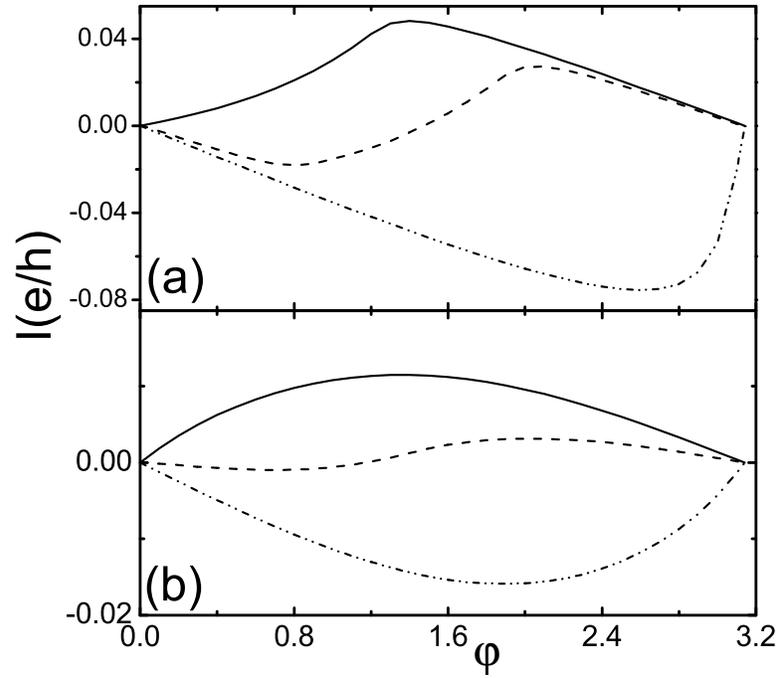}
    \caption{Current phase relation $I(\phi)$ for different barrier
strengths, (a)$t_{L}'=t_{R}'=t$ and (b) $t=_{L}'=t_{R}'=0.5t$. Other
parameters are $d=20a$, $t_{so}=0.05$, $\theta=0$ (solid line),
$\theta=0.25\pi$ (dash line), and $\theta=0.5\pi$ (dot-dash line).}
\end{figure}

\newpage

\begin{figure}[htbp]
    \includegraphics[width=4in]{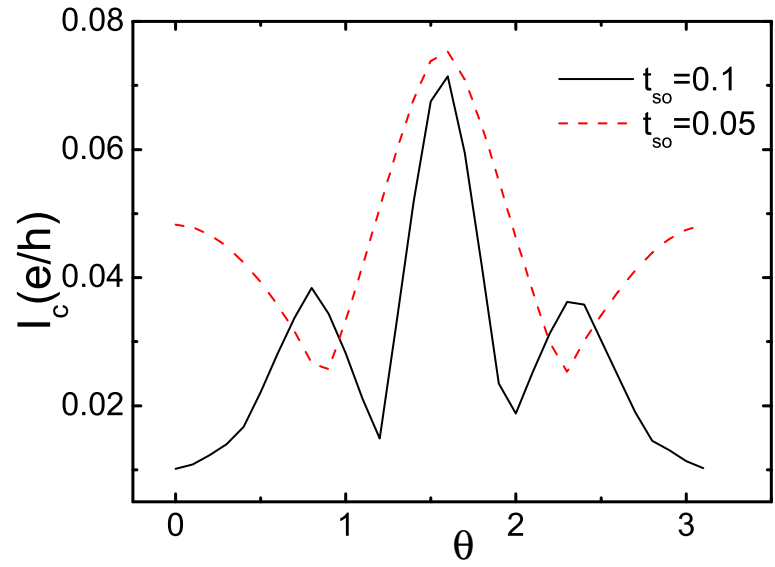}
    \caption{ The critical current as a function of angle $\theta$ for
different SOC coupling strengths, $t_{so}=0.05$ (dash line) and
$t_{so}=0.1$ (solid line), $d=20a$.}
\end{figure}

\newpage

\begin{figure}[htbp]
    \includegraphics[width=4in]{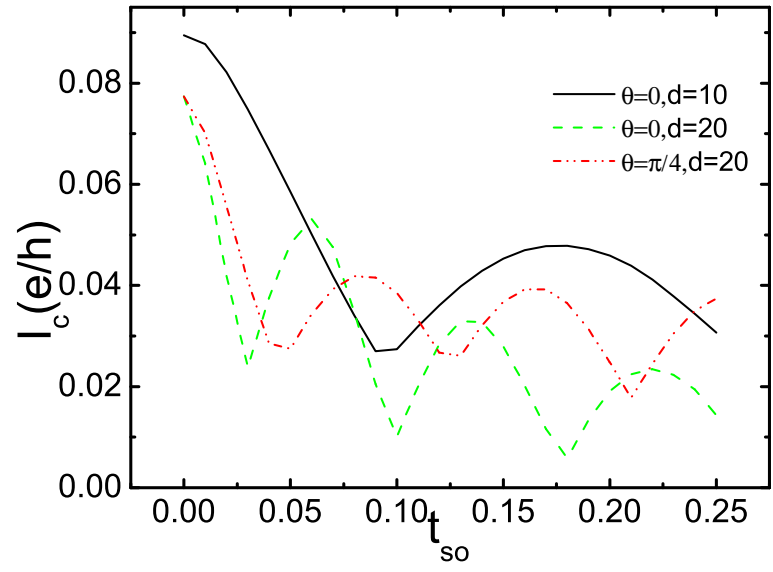}
    \caption{The critical current as a function of the SOC strength
$t_{so}$ for different 2DEG layer length $d$ and ratio angle
$\theta$.}
\end{figure}

\newpage

\begin{figure}[htbp]
    \includegraphics[width=4in]{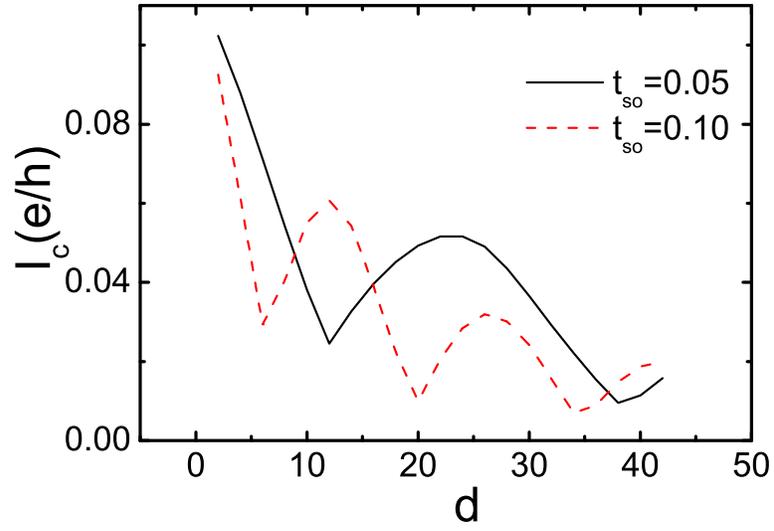}
    \caption{The critical current as a function of the 2DEG layer
length $d$ for different SOC strengths, $\theta=0$. }
\end{figure}
\end{document}